\documentclass[twocolumn]{article}
\usepackage{graphicx,epsfig}
\usepackage{amsmath}
\usepackage{amsfonts}
\usepackage{authblk}

\usepackage{ulem}
\usepackage[dvipsnames]{xcolor}
\newcommand{\tcr}{\textcolor{BrickRed}}

\begin{document}

\title{Detecting molecular folding from noise measurements}
\author[1]{Marc Rico-Pasto}
\author[2,3]{Felix Ritort}
\affil[1]{Unit of Biophysics and Bioengineering, Department of Biomedicine, School of Medicine and Health Sciences, University of Barcelona, C/Casanoves 143, 08036 Barcelona, Spain}
\affil[2]{Small Small Biosystems Lab, Condensed Matter Physics Department, Physics School, University of Barcelona, C/Mart\'i i Franqu\`es 1, 08028 Barcelona, Spain}
\affil[3]{Institut de Nanoci\'encia i Nanotecnologia (IN2UB), University of Barcelona, 08028 Barcelona, Spain}
\maketitle

\begin{abstract}
Detecting conformational transitions in molecular systems is key to understanding biological processes. Here we investigate the force variance in single-molecule pulling experiments as an indicator of molecular folding transitions. We consider cases where Brownian force fluctuations are large, masking the force rips and jumps characteristic of conformational transitions. We compare unfolding and folding data for DNA hairpin systems of loop sizes 4,8, and 20 and the 110 amino acids protein barnase, finding conditions that facilitate the detection of folding events at low forces where the signal-to-noise ratio is low. In particular, we discuss the role of temperature as a useful parameter to improve the detection of folding transitions in entropically driven processes where folding forces are temperature independent. The force variance approach might be extended to detect the elusive intermediate states in RNA and protein folding.  
\end{abstract}


\section{Introduction}
\label{intro}
Protein folding remains a challenging topic in biophysics. In 1968 Levinthal argued that stochastic diffusive motion alone could not account for the short timescales of protein folding \cite{levinthal1968}. Folding a protein into its native structure can be likened to finding a needle in a haystack. Assuming that the backbone dihedral angles of the amino acids chain are divided into three distinct regions of the Ramachandran plot, the typical folding time grows like $3^N\times\tau_d$, with $N$ the number of amino acids and $\tau_d$ the diffusive time in such regions. The latter can be expressed as $\tau_d=l^2/6D$, where $l$ is the region size, and $D$ is the diffusion constant. Taking $l\sim3$ $\rm \AA$, the inter amino-acid distance, and using the Stokes formula $D=k_BT/\gamma$ with $\gamma=6\pi\eta l$ and $\eta\sim 0.001$ Pa$\cdot$s the shear viscosity of water, we obtain $\tau_d=2\cdot 10^{-11}$ s. Thus, a protein consisting of $N=20$ residues would fold in approximately one second, while for $N=60$, the folding time would be the universe's age. This rough estimation emphasizes natural evolution's role in speeding up protein folding.  

To solve Levinthal's paradox, the molten globule hypothesis was proposed by Ptitsyn in the 70s: native folding is guided by the accumulation of native-like interactions and the sequential formation of intermediates. In small globular proteins, the molten globule is an intermediate between the unfolded and native states, where the polypeptide chain pre-forms a scaffold of the native structure. Experimental measurements suggest a dry molten globule with the outer layer of the protein hydrated and the core dehydrated. The latter has a native-like expanded structure with the backbone formed but with side chains loosely packed \cite{PTITSYN199583, vidugiris1995evidence, arai2000role}. The evidence in favor of molten globule intermediates has always been indirect \cite{semisotnov1991study, chyan1993structure, REDFIELD2004121}. 

The study of protein folding has traditionally relied on bulk experiments such as calorimetry, hydrogen exchange, NMR, and fluorescence spectroscopy. However, these methods have limitations in detecting short-lived intermediates, whose presence is masked by the averaging effect of bulk assays. Single-molecule force spectroscopy experiments have revolutionized the study of protein folding thanks to their unprecedented spatial and temporal resolution, allowing us to detect previously undetectable short-lived intermediates. Recently, using single-molecule experiments it has been demonstrated that the rupture force variance of the ligand-protein complex biotin-streptavidin increases close to the transition state \cite{cai2023anisotropic}. Optical tweezers have proven especially adept at spotting these intermediates \cite{cecconi2005, gebhardt2010, elms2012, neupane2016}, and in co-translational folding assays upon exiting the ribosome \cite{kaiser2011}. A major twist in experiments has been recently achieved with calorimetric force spectroscopy \cite{SLorenzo_2015, MRico18} by measuring the folding enthalpy, entropy, and heat capacity change of the small globular protein barnase \cite{VMitkevich_2003}. Barnase is a 110 amino acids bacterial ribonuclease protein secreted by the bacterium \textit{Bacillus amyloliquefaciens} and the focus of many studies of protein folding \cite{matouschek1989mapping, fersht1993protein, khan2003kinetic, mitkevich2003}. In reference \cite{alemany2016} we found that barnase folds in a two-state manner without observable intermediates at kHz sampling rates. In a subsequent study \cite{rico2022molten}, we demonstrated that the transition state has the thermodynamic properties of a dry molten globule: a native-like structure of high-energy and low configurational entropy relative to the native state. This study also set a thermodynamic ground on the energy landscape hypotheses (ELH) proposed by Wolynes and collaborators in the 80s. In the ELH, proteins fold along a funnel-shaped energy landscape with multiple productive folding trajectories \cite{frauenfelder1991, bryngelson1995}. 

Despite the many studies on barnase, direct observation of the hypothesized molten globule intermediate has not been possible. A major question is identifying experimental limitations to detect hidden short-lifetime states using noise force measurements. Here we address noise measurements of the unfolding and folding dynamics of barnase measured in pulling experiments at different temperatures (7-37ºC) \cite{rico2022molten}. We compare such measurements with those obtained in DNA hairpins of varying loop sizes, where the entropic barrier to folding is large, like for proteins. To this end, we measured the force variance in pulling experiments at loading rates $4-7$pN/s and $1$kHz sampling rate. We ask whether folding events can be detected in an entropy-driven process where folding forces are low, and the folding rip is indistinguishable from the noise. We also analyze the effect of decreasing temperature to reduce thermal fluctuations and increase the signal-to-noise ratio of the folding events. Detecting folding events is critical to identify folding intermediates that require additional resolution in the experiments. Here we will focus on detecting folding events in DNA hairpins and barnase, setting the basis for future studies for detecting the often elusive folding intermediates. 

\begin{figure*}[htb]
    \centering
    \includegraphics[]{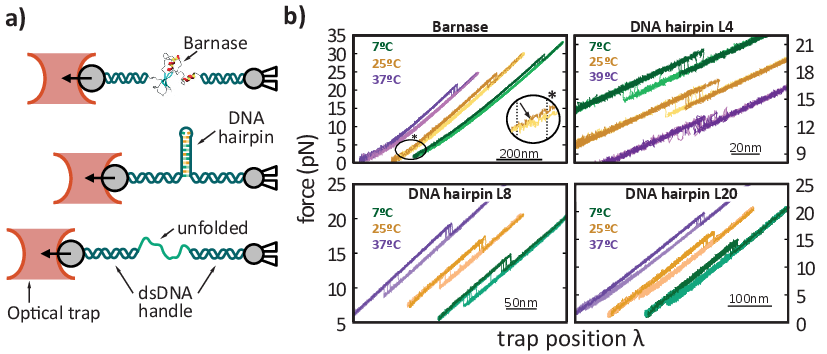}
    \caption{a) Experimental setup. Barnase and DNA hairpins are tethered between two polystyrene beads through two dsDNA handles. One bead is fixed by air suction at the tip of a micro-pipette, while the optical trap controls the other. b) Force-distance curves (FDCs) for barnase (top, left), the DNA hairpin L4 (top, right), hairpin L8 (bottom, left), and hairpin L20 (bottom, right) measured at 7ºC (green), 25ºC (orange), and 37-39ºC (purple). The dark color lines denote the unfolding trajectories, while the light color lines correspond to folding trajectories. In the starred ellipse, it is highlighted a folding event for barnase at 25ºC. }
    \label{fig1}
\end{figure*}

\section{Materials and methods}
In pulling experiments with optical tweezers, the molecule under study (DNA hairpins and barnase) is tethered between two beads. Double-stranded DNA (dsDNA) handles are attached to the end of the molecule to prevent nonspecific interactions between molecules and beads. The handles are ligated to the N- and C-termini for protein barnase via cysteine-thiol chemical reduction (details in Ref.\cite{alemany2016}). For the DNA hairpins, designed oligos are hybridized and ligated to build a DNA construct consisting of the hairpin and two flanking 29bp short handles (details in Ref. \cite{forns2011}). The 5'- end of the molecular construct is attached to one bead via anti-digoxigenin - digoxigenin bonds (3.0 to 3.4  $\mu$m diameter counts; Spherotech, Libertyville, IL), while the other end is attached to a micron-sized polystyrene microsphere using streptavidin-biotin bonds (2.0 to 2.9  $\mu$m diameter bead; G. Kisker Biotech, Steinfurt, Germany). The first bead is captured in the optical trap to measure the force, while the other is immobilized at the tip of a micro-pipette by air suction (Fig. \ref{fig1}a). 

In a pulling experiment, a molecule is tethered between two beads, and the optical trap is moved between a minimum force where the molecule is folded and a maximum force where it is unfolded. In a pulling cycle, the force applied to the system increases (decreases) when moving the optical trap away (towards) the pipette. To change the temperature, we use the temperature-jump optical trap described in Ref.\cite{SLorenzo_2015}, where an extra collimated laser is used to heat the medium surrounding the optical trap uniformly. For low-temperature measurements, the instrument is put inside an icebox kept at 4ºC, permitting us to do measurements in the range of 4-40ºC.


\section{Results and discussion}

The force is repeatedly stretched and released in pulling experiments while recording the force versus trap-position distance curves (FDCs). In the unfolding process, a force rip is observed at high forces ($>15$pN), indicating the transition from the native ($N$) to the unfolded ($U$) state (dark color trajectories in Fig.\ref{fig1}b). Furthermore, the value of force where the transition is observed varies from one pull to another, indicating that the unfolding events are thermally activated. In the refolding process, the force is reduced until a folding event is observed as a sudden force rise. The size of the force jump is proportional to the difference in molecular extension between $N$ and $U$. However, as can be seen in Fig.\ref{fig1}b (light color trajectories), a rise in the force cannot be appreciated in the folding FDCs of barnase because the folding event takes place at low forces, $<5$pN. At such low forces, the magnitude of the force jump is expected to be comparable to the noise. 

To detect the folding transition, we measured the variance of the force signal in the unfolding and folding trajectories separately. The analysis of the force variance considers the effects due to the bead, handles, and molecule under study that are modeled as three serially connected springs. The optical trap is modeled using Hooke's law,
\begin{equation} \label{eq:Stif_OT}
    f = k_b x_b
\end{equation}
where $f$ is the force, $k_b$ is the stiffness of the optical trap, and $x_b$ is the displacement of the bead to the trap's center. The dsDNA handles, and the unfolded state of the DNA hairpin and barnase are modeled with the Worm-Like Chain (WLC) model \cite{Siggia_1994},
\begin{equation}\label{eq:WLC}
    f = \frac{k_BT}{4L_p}\left( \left( 1-\frac{x}{Lc}\right)^{-2} + 4 \frac{x}{L_c} -1 \right) \, . 
\end{equation}
In Eq.\eqref{eq:WLC}, $k_B$ is the Boltzmann constant, $T$ is the temperature, $x$ is the extension of the molecule, and $L_c$ is the contour length of the handles or the unfolded molecule. Extensibility is considered for the case of the short dsDNA handles in the DNA hairpins case, by correcting the extension $x$ with the term $(1+f/Y)$ where $Y=16pN$ for the 29bp dsDNA handles \cite{forns2011}. Finally, the elastic response of the folded molecule is modeled as a dipole oriented under an applied force. Its extension is modeled with the Freely-Jointed Chain model (FJC),
\begin{equation}\label{eq:FJC}
    x = d_0\left( \coth\left(\frac{d_0 f}{k_BT}\right) - \frac{k_BT}{d_0 f}\right) \, .
\end{equation}
In Eq.\eqref{eq:FJC}, $x$ is the dipole extension at force $f$, $d_0$ is the dipole contour length, which is equal to $2$nm for the DNA hairpin, and $3$nm for barnase. 

\begin{figure*}[htb]
    \centering
    \includegraphics[]{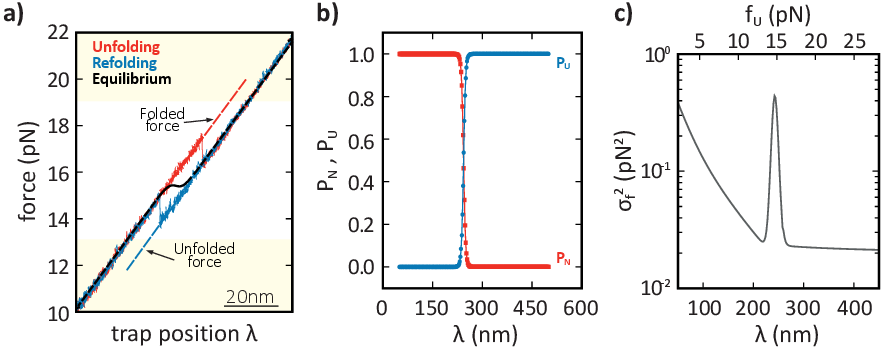}
    \caption{ Force variance calculation Eq.\eqref{VAR}. a) Unfolding (red) and folding (blue) FDCs measured for the DNA hairpin L4 at 25ºC. The red and blue dashed lines denote the folded and unfolded force branches. The black line is the calculated equilibrium FDC. b) Equilibrium probability of the native ($P_N$,red) and unfolded ($P_U$,blue) as a function of the trap position $\lambda$. c) Theoretically predicted $\sigma_f^2$ for DNA hairpin L4 at 25ºC as a function of trap position ($\lambda$,bottom x-axis) and force in the unfolded branch ($f_U$,top x-axis).  }
    \label{fig2}
\end{figure*}

\subsection{Force variance in a two-branches model}
In our pulling experiments, the control parameter is the trap position $\lambda$, and the measured force is a fluctuating quantity. To detect the folding transitions we compute the force variance ($\sigma_f^2$) in a statistical model with two branches, folded and unfolded, describing the experimental FDCs shown in Fig. \ref{fig1}. The upper and lower branches in the FDCs of Fig. \ref{fig1}b stand for the folded ($N$) and unfolded ($U$) branches where the molecule is in the Native ($N$) or Unfolded ($U$) states showing distinct FDCs. In what follows, force branches and states are used indistinctly: folded branch$\leftrightarrow$N and unfolded branch$\leftrightarrow$U. In equilibrium, the probability of observing the molecule in states $N$ or $U$ ($P_N$ and $P_U$) is given by the Boltzmann-Gibbs factor:
\begin{equation}
    P_{N(U)} = \frac{\exp\left(\frac{-\Delta G_{N(U)}}{k_BT}\right)}{Z_\lambda}
    \label{eq:Prob_NU}
\end{equation}
where $\Delta G_{N(U)}$ is the partial free energy of $N(U)$ at a given trap position and $Z_\lambda = \exp(-\Delta G_N/k_BT) + \exp(-\Delta G_U/k_BT)$ is the partition function of the system (molecule, handles, and bead). The partial free energy of the system when the molecule (DNA hairpins and barnase) is in $N$ and $U$ is calculated as:
\begin{subequations}
   \begin{align}
        \Delta G_N = & \int_0^{x_d} f_d (x')dx' +  \int_0^{x_h^N}f_h(x')dx' \nonumber \\ 
        &+ \int_0^{x_b^N}f_b(x')dx' ~ .\label{eqAGN} \\
        \Delta G_U = & \Delta G_0 + \int_0^{x_U}f_U(x')dx' +  \int_0^{x_h^U}f_h(x')dx' \nonumber \\
        &+ \int_0^{x_b^U}f_b(x')dx' ~ .\label{eqAGU}
\end{align}
\end{subequations}
where $x_d$ denotes the projected extension of the dipole, $x_U$ is the extension of the unfolded molecule, $x_h^{N(U)}$ is the extension of the handles, and $x_b^{N(U)}$ is the bead displacement, all quantities evaluated at the force when the molecule is in $N$ or $U$ (i.e., $x_b^N:= x_b(f_U)$ ; $x_b^U:= x_b(f_U)$). The forces acting on each element are defined as $f_d$ (dipole), $f_h$ (handles), $f_b$ (beads), $f_U$ (unfolded polymer) and have different elastic responses resulting in the observed different force branches of Figure \ref{fig1}b.  These relations have been defined in Eqs. \eqref{eq:Stif_OT},\eqref{eq:WLC},\eqref{eq:FJC}. Note that $f_d, f_h, f_b, f_U$ are equal at the upper integration limits in \eqref{eqAGN} and in \eqref{eqAGU}, corresponding to serially connected springs.

In the absence of force jumps between the two branches, the force variance is given by, 
\begin{equation}\label{VAR0}
    \sigma_f^2 = P_N\sigma_f^2(N)+P_U\sigma_f^2(U)
\end{equation}
The force variances in each branch, $\sigma_f^2(N)$ and $\sigma_f^2(U)$, are determined by the elastic properties of the molecular construct in that branch, $k_m(N),k_m(U)$, 
\begin{equation}\label{VAR1}
    \sigma_f^2(N,U) = \frac{k_BT k_b^2}{k_b+ k_m(N,U)}
\end{equation}
where $1/k_m(N) = 1/k_h + 1/k_d$ and $1/k_m(U) = 1/k_h + 1/k_{U}$ is the stiffness of the molecular construct, resulting from two serially connected springs of stiffnesses $k_h$ (handle) and $k_d$ (dipole for the folded state) or $k_U$ for the unfolded polymer. $k_U$ and $k_d$ are derived from Eq.\eqref{eq:WLC} and Eq.\eqref{eq:FJC}, respectively. 

At a given trap position $\lambda$, the equilibrium force $\langle f \rangle$ and its second moment $\langle f^2 \rangle$ are defined as:
\begin{subequations}
   \begin{align}
   \langle f \rangle = \frac{1}{ Z_\lambda} \left( \langle f_N \rangle e^{-\Delta G_N/k_BT} + \langle f_U \rangle e^{-\Delta G_U/k_BT} \right)\label{eqforce} \\
   \langle f^2 \rangle = \frac{1}{ Z_\lambda} \left( \langle f_N^2 \rangle e^{-\Delta G_N/k_BT} + \langle f_U^2 \rangle e^{-\Delta G_U/k_BT} \right) \label{eqf2} 
\end{align}
\end{subequations}
where $\langle f_{N(U)} \rangle$ denotes the average force when the molecule is in $N(U)$, i.e., $\langle f_{N(U)} \rangle = \partial_\lambda \Delta G_{N(U)}$ with $\Delta G_{N(U)}$ given in Eqs. \eqref{eqAGN},\eqref{eqAGU}.

To determine the variance of the force, we calculated the second derivative of the thermodynamic potential $\Delta G(\lambda)$:
\begin{subequations}
   \begin{align} \label{eq:AGL}
    &\Delta G(\lambda) = -k_BT \log(Z_\lambda) \\
    \label{eq:AGL2}
    &\langle f \rangle=\partial_\lambda \Delta G(\lambda)\\
   \label{eq:d2G_keff}
    &\partial^2_\lambda \Delta G(\lambda) = \partial_\lambda \langle f \rangle = k_{eff}
\end{align}
\end{subequations}
where $k_{eff}$ is the effective stiffness of the system along the equilibrium FDC. Using the definition of $\langle f \rangle$ Eq. \eqref{eqforce} we compute $\partial_\lambda \langle f \rangle$:
\begin{equation} \label{eq:def_dLf}
    \partial_\lambda \langle f \rangle = A(\lambda) \cdot \partial_\lambda \left( \frac{1}{Z_\lambda} \right)   +  \frac{1}{Z_\lambda} \cdot  \partial_\lambda(A(\lambda))
\end{equation}
with $A(\lambda) = \langle f_N \rangle e^{- \Delta G_N / k_BT} + \langle f_U \rangle e^{- \Delta G_U / k_BT}$. For the first term $A(\lambda) \cdot \partial_\lambda \left( 1/{Z_\lambda} \right)$, we use the definition of $Z_\lambda$ in Eqs. \eqref{eq:AGL},\eqref{eq:AGL2}:
\begin{equation} \label{eq:def_AdLZ}
    A(\lambda) \cdot \partial_\lambda \left( \frac{1}{Z_\lambda} \right)  = -\frac{A(\lambda)}{Z_\lambda} \cdot \partial_\lambda \left( \log Z_\lambda \right)=\frac{\langle f \rangle ^2}{k_BT}
\end{equation}
where we used $\langle f \rangle=A(\lambda)/Z_\lambda$. The second term, $1/Z_\lambda \cdot \partial_\lambda(A(\lambda))$, is obtained by taking the $\lambda$-derivative of the above definition for $A(\lambda)$, and using $\langle f^2 \rangle$ in Eq. \eqref{eqf2}:
\begin{eqnarray}
 \label{eq:dLf}
    &\frac{1}{Z_\lambda}\partial_\lambda A(\lambda) = \langle k \rangle - \frac{\langle f^2 \rangle+P_N\sigma_f^2(N)+P_U\sigma_f^2(U)}{k_BT}
\end{eqnarray}
where, 
\begin{equation} \label{eq:k}
    \langle k \rangle = \frac{1}{ Z_\lambda} \left( \langle k_N \rangle e^{-\Delta G_N/k_BT} + \langle k_U \rangle e^{-\Delta G_U/k_BT} \right) ,
\end{equation}
is the equilibrium stiffness and $\langle k_{N(U)} \rangle = \partial_\lambda \langle f_{N(U)} \rangle$, are the stiffnesses of each branch, equal to the slope in the corresponding force branch ($N$ or $U$). 
Introducing Eqs. \eqref{eq:def_AdLZ} and \eqref{eq:dLf} into Eq. \eqref{eq:def_dLf} and using \eqref{eq:Prob_NU},and \eqref{eq:d2G_keff}, we get: 
\begin{equation}
\label{VAR}
    \sigma_ f^2 = k_BT \left( \langle k \rangle - k_{eff} \right)+P_N\sigma_f^2(N)+P_U\sigma_f^2(U)
\end{equation}
with $\sigma_ f^2 = \langle f^2 \rangle - \langle f \rangle^2$ the force variance of the equilibrium FDC and $\sigma_ f^2(N,U)=\langle f^2(N,U) \rangle - \langle f (N,U)\rangle^2$ the variance force for each branch, Eq.\eqref{VAR1}. Notice that, $\sigma_ f^2(N,U)$ differs from $k_BT\langle k_{N(U)} \rangle$. For one branch only, e.g. $P_N=1,P_U=0$ we get $\langle k \rangle=k_{eff}=\langle k_N \rangle$ and  $\sigma_ f^2=\sigma_ f^2(N)$. In general, for systems with two branches, the slope of the FDC becomes negative in the region where the two branches coexist $P_N\sim P_U\sim 1/2$ and $k_{eff}$ can become negative (black line connecting the two branches in Fig.\ref{fig2}a).

\begin{figure*}[htb]
    \centering
    \includegraphics[width = \textwidth]{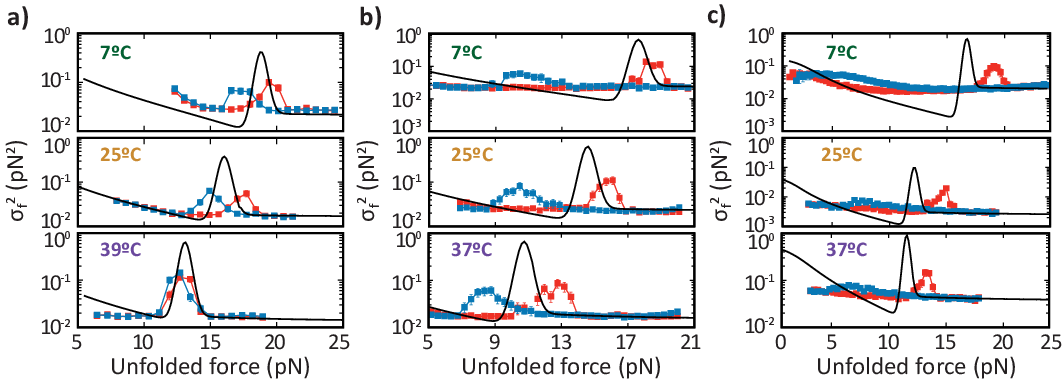}
    \caption{ Force variance $\sigma_ f^2$ for the DNA hairpin L4 (panel a), L8 (panel b), and L20 (panel c) measured at 7ºC (top), 25ºC (center), and 39ºC (bottom) as a function of the measured force along the unfolded force branch. Notice that the unfolding peak of these hairpins takes place at lower forces as we increase the temperature, while the folding peak remains independent of the temperature for L8 and L20.}
    \label{fig3}
\end{figure*}
\begin{figure*}[hbtp]
    \centering
    \includegraphics[]{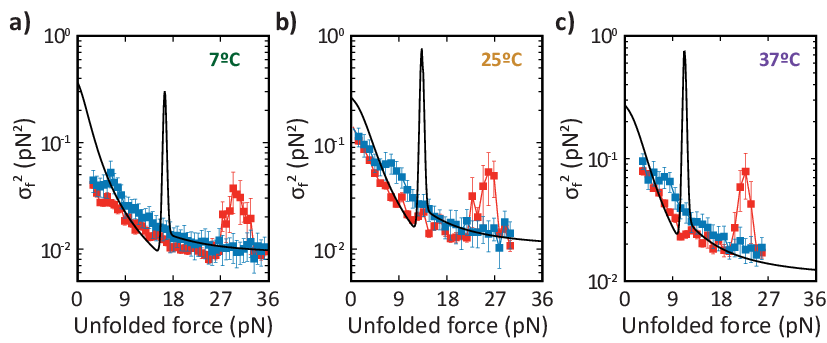}
    \caption{ Force variance $\sigma_ f^2$ for barnase measured at 7ºC (panel a), 25ºC (panel b), and 37ºC (panel c) as a function of the force along the U branch. Notice that the unfolding transition (peak in the red symbols) appears at higher forces as we decrease the temperature, while the folding event (peak in the blue symbols) does not move with temperature. }
    \label{fig4}
\end{figure*}

Figure \ref{fig2}a shows an experimental unfolding (red curve) and folding (blue curve) trajectory measured for DNA hairpin L4. Notice that at low (high) force values, $f < 13 \, (f>19)$ pN, the unfolding and folding trajectories overlap onto the folded (unfolded) branches (dashed lines), respectively. In between, unfolding and folding transitions are observed as red force rips and blue force jumps in Fig. \ref{fig2}a. To construct the equilibrium FDC (black line in Fig. \ref{fig2}a), we define the native and unfolded force branches at low and high forces outside the region limited by the force rips and jumps (red and blue dashed lines). The force branches have been calculated by fitting the elastic properties of the optical trap ($k_b$), by imposing the previously determined elastic properties of handles and unfolded polymers \cite{forns2011, rico2022molten, rico2022temp} and their folding free energies \cite{huguet2010single, rico2022molten}. This permits us to determine $\sigma_ f^2(N),\sigma_ f^2(U)$ from Eq.\eqref{VAR1} and $P_N,P_U$ from Eq.\eqref{eq:Prob_NU}. Equilibrium probabilities for each branch (red, folded; blue, unfolded) are shown in Fig. \ref{fig2}b. We derive $\sigma_f^2$ in \eqref{VAR} by computing $\langle k \rangle$ from the equilibrium FDC, and the effective stiffness of each force branch, $\langle k_N \rangle$ and $\langle k_U \rangle$. Figure \ref{fig2}c shows the estimated $\sigma_f^2$ for the DNA hairpin L4 at 25ºC as a function of the trap position (bottom axis) and the force in the unfolded branch (top axis). As expected, $\sigma_f^2$ decreases with force at low forces (F branch) and high forces (U branch) but shows a peak at the transition region $f_U\sim 15pN$ due to the contribution of the term $k_{eff}$ in \eqref{VAR}.  

The above calculations can be extended for systems with more than two branches. The average stiffness is given by:
\begin{equation} \label{k_inter}
    \langle k \rangle = (1/Z_\lambda) \sum_{m=0}^M \partial_\lambda f_m \exp(-\Delta G_m/k_BT) ,
\end{equation} 
with $M$ the total number of branches and $Z_\lambda= \sum_{m=0}^M \exp(-\Delta G_m/k_BT)$. Measuring $\langle k \rangle$ and $k_{eff}$ in equilibrium might detect intermediates by fitting the data to theoretical predictions for $M=2,3,..$. However, our pulling experiments are out of equilibrium, so the equilibrium prediction cannot be directly used to investigate the hypothesized intermediate state in barnase.

\subsection{DNA hairpins}
The experimental values of $\sigma_f^2$ for DNA hairpins were extracted from the experimental FDCs measured at loading rates of $4-6$ pN/s by averaging the force signal in $\lambda$-windows of 10nm, meaning that the force increases/decreases $\sim$ 0.5pN inside each window. Figure \ref{fig3}a shows the measured $\sigma_f^2$ along the unfolding (red) and folding (blue) process for the DNA hairpin L4 at 7ºC (top), 25ºC (center), and 37ºC (bottom) as a function of the force at the unfolded force branch, $f_U$. We remark four features from Figure \ref{fig3}a: first, the $\sigma_f^2$ values overlap at high and low forces as expected because the molecular state is the same (folded or unfolded). Second, the forces at which $\sigma_f^2$ is maximum (unfolding, red: folding, blue) shift to lower values as temperature increases. Third, the hysteresis of $\sigma_f^2$ between unfolding (red) and folding (blue) decreases with temperature. Fourth, equilibrium transitions are expected to populate forces between the two maxima. In fact, at 39ºC the measured unfolding (red) and folding (blue) $\sigma_f^2$ match the equilibrium prediction (black line) because experiments were carried out under quasi-static conditions (see Fig. \ref{fig1}b top, right). 

Regarding the DNA hairpins with loop sizes 8 and 20, we note that $\sigma_f^2$ during the unfolding process (red dots in Fig. \ref{fig3}b,c) shifts with temperature, whereas the same data during refolding (blue dots in Fig. \ref{fig3}b,c) change comparably much less with temperature. This is an indication that folding is entropically driven. Notice also that the unfolding forces where $\sigma_f^2$ is maximum (red symbols in Fig. \ref{fig3}) are similar for L4, L8, and L12, in agreement with the fact that the transition state of unfolding is located within hairpin's stem and independent of loop's size.

\subsection{Barnase}
For barnase, $\sigma_f^2$ was calculated by averaging the force over $\lambda$-windows of 8nm in the FDCs. Like for DNA hairpins, $\sigma_f^2$ during the folding process (blue points in Fig. \ref{fig4}) changes with temperature comparably much less than the unfolding process (red points in Fig. \ref{fig4}). Figure \ref{fig4} shows that barnase folds around 4 pN at the three temperatures, while the unfolding events and maximum $\sigma_f^2$ occur at 30pN at 7ºC, 26pN at 25ºC, and 22pN at 37ºC.

\section{Conclusions}
We studied the variance of the force signal, $\sigma_f^2$, in single-molecule pulling experiments. Our aim is to detect entropically driven folding at low forces where the magnitude of force fluctuations is high, and the signal-to-noise ratio of the folding events is low. Moreover, we computed the equilibrium force variance and compared it with the force variance measured in non-equilibrium conditions.

First, we studied three DNA hairpins as toy models to test the method's validity. The studied hairpins have a stem formed by 20 base pairs and four (L4), eight (L8), and twenty (L20) bases in the loop. The first studied hairpin, L4, has a small entropic barrier to folding, showing folding and unfolding transitions at sufficiently high forces (Fig. \ref{fig3}a). For L4, the force variance $\sigma_f^2$ detects the forces at which folding and unfolding transitions occur. We have also observed that the unfolding and folding transitions for L4 are temperature-dependent while the folding transitions for L8 and L20 are roughly temperature-independent indicating that the folding process is entropically driven (Figs. \ref{fig3}b,c). 

Next, we studied the folding process of protein barnase. This transition is challenging to detect in the FDCs (zoom in Fig. \ref{fig1}b), but it is observed as a gentle bump around 4pN in the force variance $\sigma_f^2$ (blue squares in Fig. \ref{fig4}). In this case, the transition is not observed as a clear maximum as in the case of DNA hairpins L4 and L8 (blue squares in Fig. \ref{fig3}a,b), because folding occurs far from equilibrium. In fact, the gentle bump observed for either L20 (blue squares in Fig. \ref{fig3}c) and barnase (blue squares in Fig. \ref{fig4}) should become a peak in equilibrium conditions (black lines), demonstrating that folding in these two molecules is highly irreversible. Indeed, hopping transitions between these molecules' folded and unfolded states cannot be observed within the experimentally accessible timescales.  

Future work should consider molecular intermediates and the usefulness of measuring the force variance $\sigma_f^2$ to detect them. Our approach might be extended by considering a theory for $\sigma_f^2$ in out-of-equilibrium conditions where detecting structural transition is challenging.

\bibliographystyle{ieeetr}
\bibliography{biblio}

\begin{thebibliography}{10}

\bibitem{levinthal1968}
C.~Levinthal, ``Are there pathways for protein folding?,'' {\em Journal de
  chimie physique}, vol.~65, pp.~44--45, 1968.

\bibitem{PTITSYN199583}
O.~Ptitsyn, ``Molten globule and protein folding,'' vol.~47 of {\em Advances in
  Protein Chemistry}, pp.~83--229, Academic Press, 1995.

\bibitem{vidugiris1995evidence}
G.~J. Vidugiris, J.~L. Markley, and C.~A. Royer, ``Evidence for a molten
  globule-like transition state in protein folding from determination of
  activation volumes,'' {\em Biochemistry}, vol.~34, no.~15, pp.~4909--4912,
  1995.

\bibitem{arai2000role}
M.~Arai and K.~Kuwajima, ``Role of the molten globule state in protein
  folding,'' {\em Advances in protein chemistry}, vol.~53, pp.~209--282, 2000.

\bibitem{semisotnov1991study}
G.~Semisotnov, N.~Rodionova, O.~Razgulyaev, V.~Uversky, A.~Gripas', and
  R.~Gilmanshin, ``Study of the “molten globule” intermediate state in
  protein folding by a hydrophobic fluorescent probe,'' {\em Biopolymers:
  Original Research on Biomolecules}, vol.~31, no.~1, pp.~119--128, 1991.

\bibitem{chyan1993structure}
C.~L. Chyan, C.~Wormald, C.~M. Dobson, P.~A. Evans, and J.~Baum, ``Structure
  and stability of the molten globule state of guinea pig. alpha.-lactalbumin:
  A hydrogen exchange study,'' {\em Biochemistry}, vol.~32, no.~21,
  pp.~5681--5691, 1993.

\bibitem{REDFIELD2004121}
C.~Redfield, ``Using nuclear magnetic resonance spectroscopy to study molten
  globule states of proteins,'' {\em Methods}, vol.~34, no.~1, pp.~121--132,
  2004.
\newblock Investigating Protein Folding, Misfolding and Nonnative States:
  Experimental and Theoretical Methods.

\bibitem{cai2023anisotropic}
W.~Cai, M.~J{\"a}ger, J.~T. Bullerjahn, T.~Hugel, S.~Wolf, and B.~N. Balzer,
  ``Anisotropic friction in a ligand-protein complex,'' {\em Nano Letters},
  2023.

\bibitem{cecconi2005}
C.~Cecconi, E.~A. Shank, C.~Bustamante, and S.~Marqusee, ``Direct observation
  of the three-state folding of a single protein molecule,'' {\em Science},
  vol.~309, no.~5743, pp.~2057--2060, 2005.

\bibitem{gebhardt2010}
J.~C.~M. Gebhardt, T.~Bornschl{\"o}gl, and M.~Rief, ``Full distance-resolved
  folding energy landscape of one single protein molecule,'' {\em Proceedings
  of the National Academy of Sciences}, vol.~107, no.~5, pp.~2013--2018, 2010.

\bibitem{elms2012}
P.~J. Elms, J.~D. Chodera, C.~Bustamante, and S.~Marqusee, ``The molten globule
  state is unusually deformable under mechanical force,'' {\em Proceedings of
  the National Academy of Sciences}, vol.~109, no.~10, pp.~3796--3801, 2012.

\bibitem{neupane2016}
K.~Neupane, A.~P. Manuel, and M.~T. Woodside, ``Protein folding trajectories
  can be described quantitatively by one-dimensional diffusion over measured
  energy landscapes,'' {\em Nature Physics}, vol.~12, no.~7, pp.~700--703,
  2016.

\bibitem{kaiser2011}
C.~M. Kaiser, D.~H. Goldman, J.~D. Chodera, I.~Tinoco, and C.~Bustamante, ``The
  ribosome modulates nascent protein folding,'' {\em Science}, vol.~334,
  no.~6063, pp.~1723--1727, 2011.

\bibitem{SLorenzo_2015}
S.~de~Lorenzo, M.~Ribezzi-Crivellari, J.~R. Arias-Gonzalez, S.~B. Smith, and
  F.~Ritort, ``A temperature-jump optical trap for single-molecule
  manipulation,'' {\em Biophysical journal}, vol.~108, no.~12, pp.~2854--2864,
  2015.

\bibitem{MRico18}
M.~Rico-Pasto, I.~Pastor, and F.~Ritort, ``Force feedback effects on single
  molecule hopping and pulling experiments,'' {\em The Journal of chemical
  physics}, vol.~148, no.~12, p.~123327, 2018.

\bibitem{VMitkevich_2003}
V.~A. Mitkevich, A.~A. Schulga, Y.~S. Ermolyuk, V.~M. Lobachov, V.~O. Chekhov,
  G.~I. Yakovlev, R.~W. Hartley, C.~N. Pace, M.~P. Kirpichnikov, and A.~A.
  Makarov, ``Thermodynamics of denaturation of complexes of barnase and binase
  with barstar,'' {\em Biophysical Chemistry}, vol.~105, no.~2-3, pp.~383--390,
  2003.

\bibitem{matouschek1989mapping}
A.~Matouschek, J.~T. Kellis~Jr, L.~Serrano, and A.~R. Fersht, ``Mapping the
  transition state and pathway of protein folding by protein engineering,''
  {\em Nature}, vol.~340, no.~6229, pp.~122--126, 1989.

\bibitem{fersht1993protein}
A.~R. Fersht, ``Protein folding and stability: the pathway of folding of
  barnase,'' {\em FEBS letters}, vol.~325, no.~1-2, pp.~5--16, 1993.

\bibitem{khan2003kinetic}
F.~Khan, J.~I. Chuang, S.~Gianni, and A.~R. Fersht, ``The kinetic pathway of
  folding of barnase,'' {\em Journal of Molecular Biology}, vol.~333, no.~1,
  pp.~169--186, 2003.

\bibitem{mitkevich2003}
V.~A. Mitkevich, A.~A. Schulga, Y.~S. Ermolyuk, V.~M. Lobachov, V.~O. Chekhov,
  G.~I. Yakovlev, R.~W. Hartley, C.~N. Pace, M.~P. Kirpichnikov, and A.~A.
  Makarov, ``Thermodynamics of denaturation of complexes of barnase and binase
  with barstar,'' {\em Biophysical chemistry}, vol.~105, no.~2-3, pp.~383--390,
  2003.

\bibitem{alemany2016}
A.~Alemany, B.~Rey-Serra, S.~Frutos, C.~Cecconi, and F.~Ritort, ``Mechanical
  folding and unfolding of protein barnase at the single-molecule level,'' {\em
  Biophysical journal}, vol.~110, no.~1, pp.~63--74, 2016.

\bibitem{rico2022molten}
M.~Rico-Pasto, A.~Zaltron, S.~J. Davis, S.~Frutos, and F.~Ritort, ``Molten
  globule--like transition state of protein barnase measured with calorimetric
  force spectroscopy,'' {\em Proceedings of the National Academy of Sciences},
  vol.~119, no.~11, p.~e2112382119, 2022.

\bibitem{frauenfelder1991}
H.~Frauenfelder, S.~G. Sligar, and P.~G. Wolynes, ``The energy landscapes and
  motions of proteins,'' {\em Science}, vol.~254, no.~5038, pp.~1598--1603,
  1991.

\bibitem{bryngelson1995}
J.~D. Bryngelson, J.~N. Onuchic, N.~D. Socci, and P.~G. Wolynes, ``Funnels,
  pathways, and the energy landscape of protein folding: a synthesis,'' {\em
  Proteins: Structure, Function, and Bioinformatics}, vol.~21, no.~3,
  pp.~167--195, 1995.

\bibitem{forns2011}
N.~Forns, S.~de~Lorenzo, M.~Manosas, K.~Hayashi, J.~M. Huguet, and F.~Ritort,
  ``Improving signal/noise resolution in single-molecule experiments using
  molecular constructs with short handles,'' {\em Biophysical journal},
  vol.~100, no.~7, pp.~1765--1774, 2011.

\bibitem{Siggia_1994}
C.~Bustamante, J.~F. Marko, E.~D. Siggia, S.~Smith, {\em et~al.}, ``Entropic
  elasticity of lambda-phage dna,'' {\em Science}, vol.~265, pp.~1599--1599,
  1994.

\bibitem{rico2022temp}
M.~Rico-Pasto and F.~Ritort, ``Temperature-dependent elastic properties of
  dna,'' {\em Biophysical Reports}, vol.~2, no.~3, 2022.

\bibitem{huguet2010single}
J.~M. Huguet, C.~V. Bizarro, N.~Forns, S.~B. Smith, C.~Bustamante, and
  F.~Ritort, ``Single-molecule derivation of salt dependent base-pair free
  energies in dna,'' {\em Proceedings of the National Academy of Sciences},
  vol.~107, no.~35, pp.~15431--15436, 2010.

\end{thebibliography}

\end{document}